\documentstyle[preprint,aps]{revtex}
\tighten
\begin{document}
\draft
\preprint{\parbox{77mm}{\flushright
        Preprint Numbers:~ANL-PHY-8072-TH-95 \\
        INT95-00-97}}
\title{Model gluon propagator and pion and rho-meson observables}
\author{M. R. Frank$^\ast $ and C. D. Roberts$^\dagger $}
\address{
$^\ast$Institute for Nuclear Theory, University of Washington, Seattle, WA
98195, USA}
\address{$^\dagger$Physics Division, Argonne National Laboratory, Argonne, IL
60439-4843, USA}
\maketitle
\begin{abstract}
A one parameter, model confined-gluon propagator is employed in a
phenomenological application of the Dyson-Schwinger and Bethe-Salpeter
equations to the calculation of a range of $\pi$- and $\rho$-meson
observables.  Good agreement is obtained with the data.  The calculated quark
propagator does not have a singularity on the real-$p^2$ axis.  A mass
formula for the pion, involving only the vacuum, dressed quark propagator, is
presented and shown to provide an accurate estimate of the mass obtained via
a direct solution of the Bethe-Salpeter equation.
\end{abstract}

\section{Introduction}
\label{introduction}
The Dyson-Schwinger equations (DSEs) provide a useful, semi-phenomenological
tool for the study of QCD.  These coupled integral equations relate the
n-point (Schwinger) functions of QCD to each other.  They provide a
nonperturbative, Poincar\'e invariant framework that enables one to correlate
hadronic observables through the properties of the Schwinger functions of the
elementary excitations in QCD; i.e., the Schwinger functions of quarks and
gluons.  (Quark and gluon propagators (2-point functions) are examples of
such Schwinger functions.)  This makes it particularly suitable for
addressing questions such as confinement and dynamical chiral symmetry
breaking and also hadronic spectroscopy and interactions.  This approach is
reviewed in Ref.~\cite{review} and has recently been applied to the study of
$\pi$-$\pi$ scattering\cite{RCSI94}, the electromagnetic pion form
factor\cite{CDRpion}, $\rho$-$\omega$ mixing\cite{rhoomega} and the anomalous
$\gamma^\ast\pi^0\rightarrow \gamma$-transition form factor.\cite{gpig}

It is possible to obtain information about such Schwinger functions via a
numerical simulation of a lattice-QCD action.\cite{Lattice,Lattice2,Lattice3}
However, in addition to the usual problems associated with identifying and
establishing the existence of the continuum limit, and recovering the global
symmetries of QCD, this also requires gauge fixing on the spacetime lattice.
Gauge fixing eliminates a number of gauge-equivalent gauge-field
configurations, thereby leading to poorer statistics.  It does not eliminate
all such configurations, however.  One is left with Gribov copies; i.e.,
gauge configurations in the gauge-fixed ensemble that are not distinct but
are related by topologically nontrivial gauge
transformations.\cite{gribov,rivers,Z94} This entails an overcounting problem
in the evaluation of gauge-fixed correlation functions.  Present studies are
encouraging, having established that this approach to the calculation of
gauge-fixed QCD Schwinger functions is feasible.\cite{Lattice2} However, the
problems identified above entail that they are currently qualitatively and
quantitatively unreliable.

Presently the most reliable estimates of the behaviour of quark and gluon
Schwinger functions are obtained in DSE studies.  The DSEs are a tower of
coupled equations and a solution is only tractable if this tower is truncated.
Truncation procedures that preserve the global symmetries of QCD are easy to
construct and implement.  This has not yet been accomplished for the local
symmetry in QCD, however, progress is being made following the realisation of
the importance of the nonperturbative structure of the fermion--gauge-boson
vertex.\cite{review,BC80,CP90,BR93,DMR94,BP94} This introduces an uncertainty
in the infrared; i.e., for $k^2< 1-2$~GeV$^2$.  However, this uncertainty is
merely quantitative.  There is general agreement on the qualitative features
of the quark and gluon 2-point Schwinger functions; i.e., 1) that the gluon
2-point function is significantly enhanced at small
spacelike-$k^2$\cite{review,BBZ,Atkin83,BP89} and that this entails an
enhancement of the momentum-dependent quark
mass-function\cite{review,MN83,PCR89,MM89,WKR91,Smek}; and 2) that for $k^2>
1-2$~GeV$^2$ the two-loop, renormalisation group improved, perturbative
results are quantitatively reliable.

Some phenomenological DSE studies have employed a parametrisation of the
2-point quark Schwinger function based on these results; for example,
Refs.~\cite{RCSI94,CDRpion,rhoomega,gpig}.  Such studies are
phenomenologically efficacious.  However, they involve the addition of new
parameters when applied to systems involving other than $u$ and $d$ quarks.

The introduction of new parameters is unnecessary when the propagator for a
quark of a given flavour is obtained directly from a quark DSE whose kernel
is determined by the 2-point gluon Schwinger function and the quark-gluon
vertex.  This procedure correlates the propagators for quarks of different
flavours via the parameters in the gluon 2-point function.  There have been
studies that employ this approach; for example Refs.~\cite{PCR89,Kugo,JM93}.
However, it is computationally more intensive and the studies therefore
addressed the calculation of a smaller class of observables.  The present
study is a first step in extending this latter approach.

Herein we employ a one parameter model gluon propagator (gluon 2-point
Schwinger function), motivated by the results of
Refs.~\cite{BBZ,Atkin83,BP89}, in a calculation of a range of $\pi$- and
$\rho$-meson observables.  The one parameter is a mass scale that can be
interpreted as marking the transition between the perturbative and
nonperturbative domains.  This model gluon propagator provides the kernel for
a quark DSE, which is solved to obtain the quark propagator (quark 2-point
Schwinger function) for real-$p^2\in (-\infty,\infty)$.  These two Schwinger
functions provide the kernel of the meson Bethe-Salpeter equation (BSE),
whose solution yields the meson mass and Bethe-Salpeter amplitude, which is a
necessary element in the calculation of decay constants and scattering
lengths, for example.  The single mass parameter determines all of these
Schwinger functions and is varied to obtain a good fit to a range of
calculated $\pi$ observables.  This illustrates the utility and economy of
the approach.

In studying the pion BSE we derive a mass formula for the pion, which
involves only the vacuum, dressed quark propagator, valid to all orders in
$m_R$, the renormalised current quark mass.  Our numerical studies show that
this formula provides an excellent estimate of the mass that is obtained by
actually solving the BSE.

The model gluon propagator is discussed in Sec.~\ref{secprop} and the quark
DSE in Sec.~\ref{secdse}.  The pion mass formula is presented in
Sec.~\ref{secmassform}.  Our numerical results are discussed in
Sec.~\ref{secres} and we summarise and conclude in Sec.~\ref{secconc}.

\section{Model gluon propagator}
\label{secprop}
In Euclidean metric\cite{Euclidean} the Landau gauge gluon propagator is
\begin{equation}
g^2 D_{\mu\nu}(k) =
\left(\delta_{\mu\nu} - \frac{k_\mu k_\nu}{k^2}\right)
\frac{g^2}{k^2[1+\Pi(k^2)]} \label{glprp1}
\end{equation}
where $\Pi(k^2)$ is the gluon vacuum polarisation.  Setting
\mbox{${\cal Z}_1^{gh} = {\cal Z}_3^{gh}$}, where ${\cal Z}_1^{gh}$ is the
renormalisation constant for the ghost-gluon vertex and ${\cal Z}_3^{gh}$
that for the ghost wave function, then
\begin{equation}
\Delta(k^2) \equiv \frac{g^2}{1+\Pi(k^2)}
\end{equation}
satisfies the same renormalisation group equation as the QCD running coupling
constant, $\alpha(k^2)$,\cite{UBG80} and hence
\begin{equation}
\label{glprp2}
\left(g^2 D_{\mu\nu}(k)\right)_R =
        \left(\delta_{\mu\nu} - \frac{k_\mu k_\nu}{k^2}\right)
                \frac{4\pi\alpha(k^2)}{k^2}~.
\end{equation}
This is sometimes described as the ``Abelian approximation'' because it
entails the QED-like Ward identity $Z_1 = Z_2$, where $Z_1$ is the
quark-gluon vertex renormalisation constant and $Z_2$ is the quark
wavefunction renormalisation constant.\cite{review}

The two-loop renormalisation group expression for the running coupling
constant only receives small ($\sim 10$\%) corrections from higher orders for
spacelike-$k^2>1$~GeV$^2$ and hence can be said to provide an accurate
representation on this domain.  For $k^2<1$~GeV$^2$, however, $\alpha(k^2)$
is not known and can only be calculated nonperturbatively.  The current
status of such studies is summarised in Ref.~\cite{review} and, as remarked
in Sec.~\ref{introduction}, gluon-DSE studies agree on the qualitative
behaviour of $\alpha(k^2)$ at small-$k^2$.  Present phenomenological
quark-DSE studies rely on an Ansatz for $\alpha(k^2<1~{\rm GeV}^2)$ motivated
by these gluon-DSE studies.

Herein we consider a parametrisation suggested by the Landau gauge studies of
Ref.~\cite{BP89}, which revealed a strong enhancement in the gluon propagator
at small spacelike-$k^2$ ($< 1$~GeV$^2$) that could be described by an
integrable singularity.  We employ the one parameter form:
\begin{equation}
\label{delk}
\Delta(k^2) = 4 \, \pi^2 \, d\,
        \left[ 4\,\pi^2\,m_t^2\,\delta^{4}(k)
                \, + \,
        \frac{1-{\rm e}^{(-k^2/[4 m_t^2])}}{k^2}\right]~,
\end{equation}
where $d=12/(33-2N_f)$, with $N_f=3$ the number of light flavours.  The first
term in Eq.~(\ref{delk}) provides an integrable, infrared
singularity\cite{MN83}, which generates long-range effects associated with
confinement, and the second ensures that the propagator has the correct large
spacelike-$k^2$ behaviour, up to $\ln[k^2]$-corrections.  A form similar to
this has been used by other authors\cite{PCR89,MM89,WKR91,Smek} with 1-loop
logarithmic corrections included in the second term.  We neglect these terms
as a simple expedient to ensure that our gluon propagator does not have a
Lehmann representation and may therefore be interpreted as describing a
confined particle; i.e., an elementary field with which there is no
associated asymptotic state.\cite{review,RWK92}

Since ours is a model gluon propagator there is no reason why the
coefficients of the two terms in Eq.~(\ref{delk}) should be related in the
particular fashion we have chosen.  However, consider
\begin{equation}
\label{delx}
\Delta(x^2)  \equiv   \int \frac{d^4k}{(2\pi )^4}
{\rm e}^{i k\cdot x} \Delta(k^2) \nonumber \\
  =  d\,\left[ m_t^2
        + \frac{1}{x^2}\,{\rm e}^{-x^2\,m_t^2} \right]~.
\end{equation}
It is clear from this that with our choice of the ratio of these coefficients
the effects of $\delta^4(k)$ in Eq.~(\ref{delk}) are completely cancelled at
small $x^2$; i.e.,
\begin{equation}
\label{delsmx}
\Delta(x^2) \stackrel{m_t^2 x^2< 1}{\simeq}
        \frac{d}{x^2} + {\rm O}(x^2)~,
\end{equation}
which is the form expected from QCD (again neglecting logarithmic-corrections).

One can therefore interpret $m_t$ as the mass scale in our model that marks
the transition from the perturbative to the nonperturbative regime.  Herein
$m_t$ is varied to provide a best fit to a range of calculated pion
observables.  [See Eq.~(\ref{params}) and the associated discussion.]

\section{quark self energy}
\label{secdse}
In Euclidean metric\cite{Euclidean} the DSE for the quark propagator is
\begin{equation}
\label{dserenorm}
S^{-1}(p)=Z_2(i\gamma \cdot p+ m_0)+\Sigma '(p),\label{s^-1}
\end{equation}
where
\begin{equation}
\label{sep}
\Sigma '(p)\equiv  Z_1\int ^{\Lambda }\frac{d^4k}{(2\pi
)^4}\,\case{4}{3}\,g^2\,
D_{\mu \nu }(p-k)\gamma _{\mu }S(k)\Gamma_{\nu }(p,k)~,
\end{equation}
with $\Gamma_\mu(p,k)$ the quark-gluon vertex, is the regularised self
energy, which can be decomposed as
\begin{equation}
\Sigma'(p) = i\,\gamma\cdot p \left(A'(p^2) - 1\right) + B'(p^2)~.
\end{equation}
The inverse of the renormalised quark propagator is
\begin{equation}
S^{-1}(p)  =  i\,\gamma\cdot p + \Sigma(p) =
        i\,\gamma\cdot p A(p^2) + B(p^2)~.
\label{sinvrenorm}
\end{equation}
Herein the prime denotes regularised quantities and unprimed quantities are
fully renormalised.

We employ a subtractive renormalisation scheme, requiring that, at a
spacelike renormalisation point, $\mu^2$,
\begin{equation}
\label{bcrenorm}
S^{-1}(p)|_{p^2=\mu ^2}=i\gamma \cdot p +m_R,
\end{equation}
with $m_R$ the renormalised {\it current } quark mass.

In this scheme, the wavefunction and mass renormalisation constants are
given by
\begin{equation}
Z_2\equiv 2 - A'(\mu ^2,\Lambda ^2) \; \; \; \mbox{and} \; \; \;
m_R\equiv Z_2\,m_0(\Lambda ^2)+B'(\mu ^2,\Lambda ^2)~, \label{wfmren}
\end{equation}
respectively, and the renormalised self energies are therefore obtained from
\begin{eqnarray}
A(p^2,\mu^2) & = & 1 + A'(p^2,\Lambda^2) - A'(\mu^2,\Lambda^2)~, \\
B(p^2,\mu^2) & = & m_R(\mu^2) + B'(p^2,\Lambda^2) - B'(\mu^2,\Lambda^2)~.
\label{rense}
\end{eqnarray}
In this scheme, $A(\mu ^2)=1$ and $B(\mu ^2)=m_R(\mu^2)$.  (In the following
we often write $m_R(\mu^2)$ as simply $m_R$, in which case the $\mu^2$
dependence is implicit.)

The renormalised axial-vector Ward identity is
\begin{equation}
\label{awi}
(p-q)_{\mu }i\,\Gamma _{\mu }^5(p,q)
        = S^{-1}(p)\,\gamma _5 +\gamma _5\,S^{-1}(q)
        - 2m_R\,\Gamma ^5(p,q)~.
\end{equation}
The composite operators $\Gamma _{\mu }^5$ and $\Gamma ^5$ are renormalised
such that, at $p^2=\mu^2=q^2$,
\mbox{$\Gamma _{\mu }^5(p,q)= \gamma_\mu\gamma_5$} and
\mbox{$\Gamma^5(p,q)=\gamma_5$}.

The chiral limit is identified as the limit in which the renormalised
axial-vector current is conserved; i.e, with the limit $m_R(\mu^2)\rightarrow
0$.

\subsection{Analysis of the large-$p^2$ behaviour of the quark propagator}
At large spacelike-$k^2$ and $p^2$ one may replace the gluon propagator and
the quark-gluon vertex by their asymptotic forms:
\begin{equation}
\Delta(k^2) \rightarrow \frac{1}{k^2}\; \; \; \mbox{and} \; \; \;
\Gamma_\mu(p,k) \rightarrow \gamma_\mu~. \label{asforms}
\end{equation}
In this limit $A(p^2)\equiv 1$ and $B(p^2)$ is the solution of
\begin{equation}
B(x) = Z_2m_0+\frac{\lambda}{4}\,\int_0^{\Lambda^2}\,dy\,y\,
        \left(\frac{1}{x}\,\theta(x-y) + \frac{1}{y}\theta(y-x)\right)\,
        \frac{B(y)}{y + B^2(y)},\label{asdse}
\end{equation}
where $x=p^2$, $y=k^2$ and $\lambda= 4\,Z_1\,d$.

For $x$ such that $B(x)^2 \ll x$; i.e., for $x \ge \mu^2$, this integral
equation is equivalent to the differential equation
\begin{equation}
\label{bdifeq}
\frac{d}{dx}\left( x^2\,\frac{d}{dx}\,B(x)\right)
        + \frac{\lambda}{4}\,B(x) = 0~,
\end{equation}
subject to the boundary condition
\begin{equation}
\label{bndconmu}
B(\mu^2) = m_R
\end{equation}
or
\begin{equation}
\left.\left( \frac{d}{dx}\left[ x B(x)\right] \right)
        \right|_{x=\Lambda^2} = Z_2\,m_0~.\label{bndconUV}
\end{equation}

Under the change of variables $x=\mu ^2{\rm exp}(2 z )$, Eq.~(\ref{asdse})
becomes
\begin{equation}
\ddot{B}(z )+2\dot{B}(z )+\lambda B(z)=0~,\label{dhoeq}
\end{equation}
which is the equation of motion for a damped harmonic oscillator.  One has
critical damping for $\lambda =\lambda _C=1$ and this yields the critical
coupling for dynamical chiral symmetry breaking; i.e., in the absence of the
first term in Eq.~(\ref{delk}), the model would still exhibit dynamical
chiral symmetry breaking for $\lambda>1$.  This behaviour has been observed
in QED\cite{FK76} and phenomenological models of QCD without an
infrared-singular model gluon propagator.\cite{H84,AJ88,RM90}

The solution of Eq.~(\ref{bdifeq}) consistent with Eq.~(\ref{bndconmu}) is
\begin{equation}
\label{basympt}
B(z) = \kappa\,{\rm e}^{-z}\,\cos\left(z\,\sqrt{\lambda-1} + \phi\right)~,
\end{equation}
with
\begin{equation}
\kappa\,\cos\phi = m_R~.
\end{equation}
In the chiral limit $m_R=0$ and hence $\phi=\pi/2$.  In general $\kappa$ is
only
determined in a complete solution of the integral equation.

The boundary conditions in Eqs.~(\ref{bndconmu}) and (\ref{bndconUV}) are
equivalent: a given value of $m_R$ entails a given value of $Z_2\,m_0$ and
vice-versa.  In fact, for finite $\Lambda$, $m_R=0$ generally entails
$Z_2\,m_0\neq 0$.  It follows from Eq.~(\ref{basympt}), however, that for any
finite value of $m_R$
\begin{equation}
\lim_{\Lambda^2\rightarrow \infty}\,Z_2(\mu^2,\Lambda^2)\,m_0(\Lambda^2)
                = 0~.
\end{equation}

Equation~(\ref{basympt}) indicates that the renormalised mass function will
exhibit damped oscillations about zero for $p^2>\mu^2$, a feature we observed
in our numerical solutions, which were well described by Eq.~(\ref{basympt})
on $p^2\in [\mu^2,\Lambda^2]$.  With the exception of Ref.~\cite{HW95}, other
DSE studies implicitly use $\mu=\Lambda$ and hence the oscillations are not
observed.  The oscillations were observed in Ref.~\cite{HW95}, which
addresses in detail the nonperturbative renormalisation of the fermion DSE in
QED.

\subsection{Additional remarks}
The ``Abelian approximation'' entails that $Z_1 = Z_2$ in
Eqs.~(\ref{dserenorm}) and (\ref{sep}).  We make this identification
hereafter.

In the numerical studies described below we employed the rainbow
approximation:
\begin{equation}
\label{rainbow}
\Gamma_\mu(p,k)=\gamma_\mu~.
\end{equation}
This is a quantitatively reliable approximation in Landau gauge.  (This is
not the case in other gauges).  For example, in studies of the critical
coupling for dynamical chiral symmetry breaking, a comparison of the results
obtained using this approximation\cite{FK76} with those obtained using more
realistic vertex Ans\"atze\cite{review,BC80,DMR94,BP94,Atkin93} shows this
approximation to be accurate to 5\%.  The improvements to this approximation
are qualitatively important\cite{review,BC80,DMR94,BP94,Atkin93}, being
crucial to the restoration of multiplicative renormalisability and gauge
covariance. However, herein a quantitatively reliable calculation scheme is
sufficient and this is provided by Eq.~(\ref{rainbow}) in Landau gauge.

\section{A Pion Mass Formula}
\label{secmassform}
The unrenormalised BSE for the pion in generalised-ladder approximation is,
with unrenormalised n-point functions beyond denoted by $\tilde{\cdot}$,
\begin{equation}
\label{bse}
\tilde{\Gamma}_\pi(p;P) + \int\,\frac{d^4q}{(2\pi )^4}\,
        \case{4}{3}\,g^2\,\tilde{D}_{\mu\nu}(p-q)\,\gamma_\mu\,
        \tilde{S}(q+\case{1}{2}P)\,\tilde{\Gamma}_\pi(q;P)\,
        \tilde{S}(q-\case{1}{2}P)\,
        \gamma_\nu  = 0~,
\end{equation}
where $P=p_1+p_2$ is the total momentum and $p=(p_1-p_2)/2$ the relative
momentum of the $\bar q$-$q$ pair.

For the pion it is a good approximation\cite{JM93,sepetal} to write
\begin{equation}
\label{pileading}
\tilde{\Gamma}_\pi(p;P) = \gamma_5 \tilde{F}(p^2,P^2)~,
\end{equation}
in the sense that $\tilde{\Gamma}_\pi(p;P)$ is a general pseudoscalar
$4\times 4$ matrix and the right-hand-side is, pointwise, a good
approximation to it and the inclusion of the other allowed Dirac amplitudes
alters the mass eigenvalue by $< 1$~\%.  With this approximation
Eq.~(\ref{bse}) becomes~$[C_2(R) = (N_c^2-1)/(2N_c) = 4/3~{\rm for}~N_c=3]$
\begin{equation}
\label{bseconv}
8\,N_c\,\tilde{F}(p^2,P^2) = 3\,C_2(R)\,\int\,\frac{d^4q}{(2\pi )^4}\,
        \tilde{\Delta}(p-q)\,\tilde{H}(q;P)
\end{equation}
with
\begin{equation}
\tilde{H}(p;P) = 8\,N_c\,\left(p_+ \cdot p_
        - \tilde{\sigma}_V^+ \tilde{\sigma}_V^-
                + \tilde{\sigma}_S^+ \tilde{\sigma}_S^- \right)\,
                \tilde{F}(p^2,P^2)~,
\end{equation}
where we have defined $p_\pm = p \pm P/2$,
\begin{eqnarray}
\tilde{\sigma}_V^\pm = \frac{\tilde{A}(p_\pm^2)}
        {p_\pm^2\,\tilde{A}(p_\pm^2)^2 + \tilde{B}(p_\pm^2)^2}
\;\; & \mbox{and} & \;\;
\tilde{\sigma}_S^\pm = \frac{\tilde{B}(p_\pm^2)}
        {p_\pm^2\,\tilde{A}(p_\pm^2)^2 + \tilde{B}(p_\pm^2)^2}~.
\end{eqnarray}

Equation~(\ref{bseconv}) is a convolution in four dimensions and can be
rewritten as
\begin{equation}
\label{bse3}
0= 8\,N_c\,\tilde{F}_P(x) -  C_2(R)\,3\,\tilde{\Delta}(x)\,\tilde{H}_P(x)
\end{equation}
with $\tilde{H}_P(x)$ the Fourier transform, with respect to $p$, of
$\tilde{H}(p;P)$.

Multiplying the right-hand-side of Eq.~(\ref{bse3}) by
$(\tilde{F}_P(-x)/[3 C_2(R)\tilde{\Delta}(x)])$ one can construct
\begin{equation}
\Pi_\pi(P) \equiv \int\,d^4x\,\left( \frac{8 N_c}{3 C_2(R)}
        \frac{\tilde{F}_P(-x)\,\tilde{F}_P(x)}{\tilde{\Delta}(x)}
        - \tilde{F}_P(-x)\,\tilde{H}_P(x) \right)~.
\end{equation}
In the auxiliary-field bosonisation of the Global Colour-symmetry
Model\cite{review,CR85} the effective action contains the term
\begin{equation}
\int d^4x\,d^4y\,\pi^i(x)\,\Pi_\pi(x-y)\,\pi^i(y)~,
\end{equation}
with $\pi^i(x)$ a local field variable identified with the pion field.  One
sees from this that $\Pi_\pi(P)$ plays the role of the inverse propagator for
the composite pion field.  Further, at the solution of the BSE,
$P^2=-m_\pi^2$, Eq.~(\ref{bse3}) is satisfied and hence
\begin{equation}
\Pi_\pi(P^2= - m_\pi^2) = 0~.
\end{equation}

It has been shown\cite{DS79} that for $m_0=0$ the unrenormalised
BSE has a massless, $P^2=0$, solution with
\begin{equation}
\label{Goldthma}
\tilde{F}_P(x) = \tilde{F}_{P=0}(x) = \tilde{B}_{m_0=0}(x)~,
\end{equation}
which is the manifestation of Goldstone's theorem in the DSE approach.  Using
this as an approximation for $P^2= -m_\pi^2\neq 0$, via the unrenormalised DSE:
\begin{equation}
\tilde{B}_{m_0=0}(x)= 3\,C_2(R)\,
        \tilde{\Delta}(x)\,\tilde{\sigma}_S^{m_0=0}(x)~,
\end{equation}
one obtains
\begin{equation}
\Pi_\pi(P) \approx \int\,d^4x\,\tilde{B}_{m_0=0}(x)\,\left(
        8 N_c\,\tilde{\sigma}_S^{m_0=0}(x) - \tilde{H}_P(x) \right)
        \equiv \bar \Pi_\pi(P)~.
\end{equation}

This is manifestly invariant under renormalisation and hence one may write
\begin{equation}
\label{Pirenorm}
\bar\Pi_\pi(P) = \int\,d^4x\,B_{m_R=0}(x)\left(
        8 N_c\,\sigma_S^{m_R=0}(x)  - H_P(x;m_R) \right)~,
\end{equation}
with every quantity on the right-hand-side renormalised ($\sigma_S$ and $H$
have the same form but with unrenormalised quantities replaced by
renormalised ones) and evaluated with $m_R\neq 0$ unless otherwise specified.

As remarked above, $\Pi(P^2=-m_\pi^2)=0$ at the solution of the BSE.
Equation~(\ref{Pirenorm}) therefore allows one to obtain a simple pion mass
formula derived from the generalised-ladder approximation to the BSE and
expressed solely in terms of the massless and massive renormalised, vacuum,
dressed quark propagators.

For the pion (because $m_\pi^2 \simeq 0$) it is a good approximation to write
\begin{equation}
\label{massform}
\bar\Pi_\pi(P) \approx \bar\Pi_\pi(0) + P^2\,N_\pi^2
\end{equation}
where
\begin{eqnarray}
\lefteqn{N_\pi^2 = \left(\frac{d}{dP^2}\bar\Pi_\pi(P^2)\right)_{P^2=0}=}\\
& &
 \frac{N_c}{8\pi^2}\int_0^{\Lambda^2}\,ds\,s\,B_{m_R=0}(s)^2\,
\left(  \sigma_{V}^2 - 2 \left[\sigma_S\sigma_S' + s
\sigma_{V}\sigma_{V}'\right]
  - s \left[\sigma_S\sigma_S''- \left(\sigma_S'\right)^2\right]
- s^2 \left[\sigma_V\sigma_V''- \left(\sigma_V'\right)^2\right]\right)~,
\nonumber
\end{eqnarray}
with the primes denoting differentiation with respect to $s=p^2$ and
$\sigma_V$ and $\sigma_S$ evaluated at $m_R$.  This is just the conventional,
generalised-ladder approximation Bethe-Salpeter amplitude normalisation
constant, calculated neglecting small ($\sim$ 2\%) O$(m_\pi^2)$ corrections.

We note that if $A(p^2)\equiv 1$, $N_\pi=f_\pi$.  In general, the
approximation $N_\pi\approx f_\pi$ is accurate to within 10\% and the
difference is a measure of the error introduced by the approximation of
Eq.~(\protect\ref{pileading}).\cite{review} (Also see Table.~\ref{tabobs}.)

Equation~(\ref{massform}) yields the explicit pion mass formula\cite{RTC}
\begin{equation}
\label{expmass}
m_\pi^2\,N_\pi^2 =
\frac{N_c}{2\pi^2}\,\int_0^{\Lambda^2}\,ds\,s\,
        \frac{B_{m_R=0}(s)}{B_{m_R\neq0}(s)}
    \left(B_{m_R\neq0}(s)\,\sigma_S^{m_R=0}(s)
        - B_{m_R=0}(s)\,\sigma_S^{m_R\neq 0}(s)\right)~.
\end{equation}

One notes immediately that, for a given value of $m_R$, $m_\pi^2 \rightarrow
\mbox{constant} < \infty$ as $N_c\rightarrow \infty$ and that, for arbitrary
$N_c$, $m_\pi^2\rightarrow 0$ as \mbox{$m_R\rightarrow 0$}.  Further, if the
DSE is solved with a quark-gluon vertex that ensures multiplicative
renormalisability then $m_\pi^2$ is a renormalisation point invariant and the
result is independent of the cutoff $\Lambda^2$.  The integral on the
right-hand-side of Eq.(\ref{expmass}) is convergent in the limit
$\Lambda^2\rightarrow\infty$.

{}From Eq.~(\ref{expmass}) one can recover what is sometimes called the
Gell-Mann-Oakes-Renner relation in the form:
\begin{equation}
\label{GMOR}
m_\pi^2\,f_\pi^2 =
        - \,m_R^{\mu^2} \, \langle \bar q q \rangle_{\rm vac}^{\mu^2}~,
\end{equation}
where
\begin{equation}
\label{convqbq}
-\langle \bar q q \rangle_{\rm vac}^{\mu^2}
= \frac{N_c}{2\pi^2}\,\int_0^{\Lambda^2}\,ds\,s\,\sigma_S^{m_R=0}(s)~,
\end{equation}
which is the customary definition of the vacuum condensate.  However, in
terms of the nonperturbatively dressed quark propagator, equality between the
integrands requires the following {\it ad hoc} and mutually incompatible
``approximations'': $\forall s$,
\begin{mathletters}
\begin{equation}
\label{approxa}
B_{m_R=0}(s) \approx B_{m_R\neq0}(s)~;
\end{equation}
\begin{equation}
\sigma_S^{m_R=0}(s) \approx\sigma_S^{m_R\neq 0}(s)~;
\end{equation}
\begin{equation}
\label{approxc}
B_{m_R\neq0}(s)\approx m_R + B_{m_R=0}(s)~,
\end{equation}
\end{mathletters}
which yields Eq.~(\ref{GMOR}) when one makes the additional approximation
$N_\pi\approx f_\pi$, discussed above.  That these are bad ``approximations''
is clear; for example, Eq.~(\ref{approxa}) has the effect of replacing a
convergence factor in the integrand by unity and it is incompatible with
Eq.~(\ref{approxc}).  As elucidated in Ref.~\cite{RCP88}, Eq.~(\ref{GMOR})
can only be obtained if the (renormalised) current quark mass is treated
strictly as a perturbation.  The inadequacy of Eqs.~(\ref{GMOR}) and
(\ref{convqbq}) is only exposed by a careful treatment of the Dyson-Schwinger
and Bethe-Salpeter equations.

We emphasise that Eq.~(\ref{expmass}) is completely consistent with the
general arguments of Ref.~\cite{GMOR}.  It is derived from the generalised
ladder BSE and measures the expectation value of the explicit chiral symmetry
breaking term in the pion state under the approximation that
Eq.~(\ref{Goldthma}) is valid for $P^2\neq 0$, which is why the
right-hand-side involves only vacuum quantities: massless and massive,
renormalised, vacuum, dressed quark propagators.

We demonstrate below that Eq.~(\ref{expmass}) provides an extremely accurate
estimate of the pion mass obtained by solving the pion BSE in
generalised-ladder approximation.  (See Eq.~(\ref{masslinear}) and
Table~\ref{tabobs}.)

\subsection{Solving the pion Bethe-Salpeter equation.}
\label{secbse}
In our numerical studies we are interested in the subtractively renormalised
Bethe-Salpeter amplitude, $F(p;P)$.  This is defined in terms of the
regularised
amplitude $F'(p;P)$ via
\begin{equation}
F(p;P) \equiv F'(p;P)-F'(\mu,P)~,
\end{equation}
which, in generalised ladder approximation, is obtained as the solution of
\begin{eqnarray}
\label{bsepi}
F'(p;P) & = &
Z_2\,3\,C_2(R)\,\int^\Lambda\,\bar d^4q\,\Delta(p-q)\,
\left(q_+ \cdot q_- \,\sigma_V^+ \sigma_V^-
                + \sigma_S^+ \sigma_S^- \right)\,F(q;P)~.
\end{eqnarray}
It is clear that all corrections to free-field behaviour vanish at the
renormalisation point; i.e., $\left.F(p;P)\right|_{p^2=\mu^2}=0$.

Upon comparison with the DSE for $B(p^2)$ in
Sec.~\ref{secdse}, it is clear that in the chiral limit $(m_R=0)$ one has
\begin{equation}
\label{goldthm}
F(p;P)=B_{m_R=0}(p)~;
\end{equation}
i.e., that Goldstone's theorem is manifest.\cite{DS79}

One may solve Eq.~(\ref{bsepi}) numerically by introducing an eigenvalue,
$\lambda(P^2)$, on the right-hand-side.  This yields an equation that has a
solution at every value of $P^2$.  The equation can then be solved repeatedly
until that $P^2$ is found for which $\lambda(P^2)=1$.

The eigenvalue and eigenvector are determined by employing the Tschebyshev
decomposition
\begin{equation}
F(p;P) = \sum_{i=1}^\infty\,F_i(p^2,P^2)\,U_i(\cos\beta)
\end{equation}
and solving for the Tschebyshev moments of $F(p;P)$, which are obtained via
\begin{equation}
F_i(p^2,P^2) = \case{2}{\pi}\,\int_0^\pi\,d\beta\,\sin^2\beta\,
                U_i(\cos\beta)\,F(p,P)~.
\end{equation}
In practice we only keep the lowest moment $F_0(p^2,P^2)$; neglecting the
coupling to the higher moments.  This is a very good approximation for the
pion.\cite{JM93}

For an on-shell pion $P^2=-m_\pi^2$ and hence the right-hand-side of
Eq.~(\ref{bsepi}) samples the quark propagator at complex values of its
argument.  To avoid solving the quark DSE off the real-$p^2$ axis we expanded
\mbox{$(q_+\cdot q_-\, \sigma_V^+ \sigma_V^-  + \sigma_S^+ \sigma_S^-)$}
to O$(P^2)$ and solved the resulting equation, which involves derivatives of
the propagator at real-$p^2\geq 0$.

\section{Numerical Results and Phenomenology}
\label{secres}
We have two parameters: the mass scale $m_t$ in the gluon propagator, which
marks the transition point between the perturbative and nonperturbative
domains, Eq.~(\ref{delsmx}); and $m_R$, the renormalised current quark mass.
We varied these parameters in order to obtain the best $\chi^2$-fit to the
pion observables: $m_\pi$ [calculated using Eq.~(\ref{expmass})], the weak
pion decay constant\cite{review}
\begin{equation}
f_\pi = \frac{N_c}{4\pi^2}\int_0^{\Lambda^2}\,
        ds\,s\,\case{1}{N_\pi}\,F_0(s,P^2)\,
        \left[\sigma_V\sigma_S
        + \case{1}{2}s\left(\sigma_V'\sigma_S
                - \sigma_V\sigma_S'\right)\right]~,
\end{equation}
$r_\pi$ and the $\pi$-$\pi$ scattering lengths: $a_0^0$, $a_0^2$, $a_1^1$,
$a_2^0$, expressions for which are given in Ref.~\cite{RCSI94}.

At each pair of parameter values the quark DSE was
solved numerically with $\mu = 48$~fm$^{-1}=9.47$~GeV, which is large enough
to be in the purely perturbative domain, and $\Lambda= 2^{18}$~fm$^{-1}\sim
5461\mu$.  The results were almost independent of the cutoff; doubling it
leading only to a 3\% change in $f_\pi$, for example.  Our results would have
been completely independent of $\Lambda$ if we had employed a vertex that
preserves multiplicative renormalisability.  This observation provides a
quantitative measure of the violation of multiplicative renormalisability
when the rainbow approximation is used in Landau gauge.  It is significantly
worse in other gauges.  As remarked above, rainbow approximation entails a
loss of gauge covariance.  Our experience suggests that our results would
change by no more than 10\% if we had used a dressed fermion--gauge-boson
vertex that ensured gauge covariance of the fermion
propagator.\cite{review,DMR94,BP94,Atkin93}

The formulae for the observables were then evaluated using the solution
obtained and the approximation that Eq.~(\ref{goldthm}) is valid for $m_R\neq
0$.  After obtaining the optimal values of the parameters we recalculated the
observables using the pion Bethe-Salpeter amplitude calculated as described
in Sec.~\ref{secbse}.  We found numerically that
\begin{equation}
\label{feqb}
F_0(p^2;P^2) \approx B_{m_R=0}(p^2)~.
\end{equation}

The best $\chi^2$-fit was obtained with
\begin{equation}
\label{params}
\begin{array}{lcl}
m_t = 0.69~\mbox{GeV} &\;\;\mbox{and}\;\; & m_R = 1.1~\mbox{MeV}~.
\end{array}
\end{equation}
We also carried out an extended $\chi^2$-fit where the ratio of the
coefficients of the two terms in Eq.~(\ref{delk}) was allowed to vary.  In
this case the best $\chi^2$ was obtained with the value of $m_t$ in
Eq.~(\ref{params}) and a ratio that agreed with that in Eq.~(\ref{delk}) to
within 2\%.  The data therefore requires both terms in the propagator and the
cancellation of long-range effects described in Eq.~(\ref{delsmx}).

The observables calculated with these parameter values are presented in
Table~\ref{tabobs}.  One observes immediately that our one parameter model
for the gluon propagator provides a good description of low energy pion
observables.  This improves upon the results of
Refs.~\cite{RCSI94,CDRpion,rhoomega,gpig}, in which the quark propagator was
parametrised and illustrates the connection, suggested in these articles,
that may be made between hadronic observables and the quark-quark
interaction.

We have made a direct comparison on the spacelike-$p^2$ axis of the numerical
solutions for $\sigma_V$ and $\sigma_S$ obtained herein with the parametrised
forms used in Ref.~\cite{CDRpion}.  The agreement in form and magnitude is
very good, which suggests that the one parameter model gluon propagator will
also provide a good description of hadronic form factors.

One observes that the mass formula in Eq.~(\ref{expmass}) yields an accurate
estimate of the mass obtained by solving the pion BSE.  We find that, with
parameters of Eq.~(\ref{params}), the right-hand-side of Eq.~(\ref{expmass})
is well described by
\begin{equation}
\label{masslinear}
m_\pi^2 N_\pi^2 = 2\,(0.45)^3\,m_R + (2.6)^2\,m_R^2 + 150\, m_R^3
\end{equation}
on the range $m_R\in [0,0.02]$~GeV, from which one may infer a value of
\mbox{$\langle \bar q q\rangle_{\mu} = -(0.45~{\rm GeV})^3$}.  At the value of
$m_R$ in Eq.~(\ref{params}) the term linear in $m_R$ contributes almost 96\%
of the total.  We see, therefore, that Eq.~(\ref{expmass}) entails $m_\pi^2
\propto m_R$, for small $m_R$, but that the constant of proportionality is not
given by the usual definition of the vacuum quark condensate,
Eq.~(\ref{convqbq}).

Our one parameter model for the gluon propagator explicitly {\it excludes}
the $\ln[k^2]$-corrections associated with the anomalous dimensions in QCD.
It is therefore inappropriate to directly compare $m_R(\mu)$ in
Eq.~(\ref{params}) with the QCD evolution of the commonly quoted value of
$m_{\mu=1 {\rm GeV}}\approx 7.5~{\rm MeV} $.~\cite{PDG94} (This entails that
the same is true of \mbox{$\langle \bar q q\rangle_{\mu}$}.)  We note that
replacing $(\pi d)/k^2$ by $\alpha_S^{\rm two-loop}(k^2)/k^2$ in
Eq.~(\ref{delk}) would lead to a suppression of the tail of the quark mass
function, thereby requiring a larger value of $m_R$ to reproduce the pion
mass and a commensurate change in $m_t$.  This represents a quantitative
improvement of our model but would not change its qualitative features.

\subsection{$\rho$-meson observables.}
\label{secrho}
We have employed our model gluon propagator in a preliminary study of
$\rho$-meson properties.

The regularised, generalised ladder approximation to the $\rho$-meson BSE is
\begin{equation}
\label{bserho}
F_\rho'(p;P) =  Z_2\,3\,C_2(R)\,\int^\Lambda\,g^2\,D_{\mu\nu}(p-q)\,
        \case{1}{12}{\rm tr}\left[\gamma_\alpha i\gamma_\mu\,S(q_+)\,
        iT_\alpha(P)\,S(q_-)\gamma_\nu\right]\,F_\rho(q;P)~,
\end{equation}
where $[T_\mu(P) = \gamma_\mu + \gamma\cdot P P_\mu/m_\rho^2]$.  The
subtractively renormalised amplitude is given by
\mbox{$F_\rho(p;P)=F_\rho'(p;P) - F_\rho'(\mu;P)$}.  We neglected
the other Dirac-structures allowed in the vector-meson Bethe-Salpeter
amplitude.  For the $\rho$-meson the error introduced by this truncation is
approximately 10\%.\cite{sepetal} The $\rho$- and $\omega$-mesons are
degenerate at this level of approximation.  As for the pion, we project this
equation onto the lowest Tschebyshev moment and solve for $F_0(p^2,P^2)$,
neglecting the coupling to the higher moments.  This is a good approximation
for the $\rho$-meson.\cite{JM93}

In this preliminary study we have only solved the quark DSE at real-$p^2$.
For an on-shell $\rho$-meson $P^2<0$ and hence Eq.~(\ref{bserho}) samples the
quark propagator at complex values of $p^2$.  To obtain an approximate
solution of Eq.~(\ref{bserho}), without solving the quark DSE at
complex-$p^2$, we introduced an eigenvalue, $\lambda(P^2)$, on the
right-hand-side of Eq.~(\ref{bserho}) and solved this equation at spacelike
values of $P^2$, thereby obtaining $\lambda(P^2>0)$.  For $0<P^2<10~{\rm
fm}^{-2}$ the results could be described by the quadratic (in $P^2$):
\begin{equation}
\lambda(P^2) = 0.44 - 0.021\,P^2 + 0.000076\,P^4
\end{equation}
with a standard-deviation of $0.000044$.  We compared this with both linear
and cubic fitting forms: it provides a smaller standard-deviation than the
linear form and is monotonic, whereas the cubic is not.  The value of $P^2$
for which this algebraic form of $\lambda(P^2)=1$ provides the mass estimate
presented in Table~\ref{tabobs}.

The calculated $\rho$-meson Bethe-Salpeter amplitude is much narrower in
momentum space than that of the pion, in agreement with the results of
Ref.\cite{JM93}.

The calculation of $g_{\rho\pi\pi}$ proceeds in a similar manner.  In
generalised impulse approximation the $\rho\pi\pi$ coupling can be expressed
in terms of a nonlocal coupling functional, $N_\mu(p,q)$, which is discussed
in Ref.~\cite{PRC87}.  This expression is used to evaluate
$g_{\rho\pi\pi}(P^2)$ on $0<P^2<10~{\rm fm}^{-2}$.  The results were fitted
and extrapolated to the calculated mass-shell point.  The best fit was
obtained with:
\begin{equation}
g_{\rho\pi\pi}(P^2)= 1.15 - 0.076\,P^2 + 0.0013\,P^4 - 0.000022 P^6~,
\end{equation}
giving a standard-deviation of 0.00023.  This form provides a smaller value
of the standard-deviation than either a linear or quadratic form and is
monotonic whereas the quartic is not.  The value obtained at the calculated
on-mass-shell point is given in Table~\ref{tabobs}.

These calculations are only a first step.  They serve merely to indicate that
our one parameter model gluon propagator, which was fitted to pion
observables, can reasonably be expected to provide a good description of
other observables too.

\subsection{Confinement.}
We have also solved the quark DSE for real-$p^2<0$.  There is no singularity
on the real-$p^2$ axis.  The solution therefore does not have a Lehmann
representation and hence may be interpreted as describing a confined
particle.

A plot of $1/[p^2 + M(p^2)^2]$, which for a free particle would have a pole
at the mass-shell point, has a broad resonance-like peak centred on
$p^2\approx -(0.55)$~GeV$^2$.  This admits an interpretation as the
``constituent-quark-mass'' in our model.

The form of our solution is suggestive of a pair of complex conjugate poles
or branch points with timelike real parts and large magnitude imaginary
parts.  We have made no attempt to confirm this.  A thorough study must
identify whether this structure is an artifact of the rainbow approximation,
which is known to be associated with unexpected behaviour of the fermion
propagator in the complex plane\cite{AB79,MH92,SC92,Maris,Stainsby} that is
modified when the vertex is dressed.\cite{BRW92}

\section{Summary and conclusions}
\label{secconc}
Using a confining, one parameter model form for the gluon propagator,
Eq.~(\ref{delk}), which incorporates the essence of the solution of
realistic, approximate gluon Dyson-Schwinger equations (DSEs), we solved the
renormalised, rainbow approximation quark DSE and subsequently the
renormalised, generalised ladder approximation $\pi$- and $\rho$-meson
Bethe-Salpeter equations (BSEs).  We varied the parameter in the gluon
propagator, $m_t$, which is a mass scale that marks the transition between
the perturbative and nonperturbative domains, and the renormalised current
quark mass and obtained a good description of a range of $\pi$- and
$\rho$-meson observables.  The value of $m_t$ was not known {\it a priori}.
Good agreement with the data {\em required} $m_t\sim 700$~MeV, which
corresponds to a length of $\sim 0.3$~fm.

In studying the pion BSE we were led to a mass formula
for the pion, Eq.~(\ref{expmass}), expressed solely in terms of the massive
and massless quark propagators.  This formula provides a very accurate
estimate of the pion mass.  It is valid to all orders in $m_R$, the
renormalised current quark mass, and for $m_R<20$~MeV the nonlinear terms
provide a contribution of no more than $\sim$10\%.

We obtained numerical solutions of the quark DSE on the timelike-$p^2$ axis,
which showed the quark propagator to have no singularity on the real-$p^2$
axis in our model.  We found evidence to suggest that, as a function of
$p^2$, the quark propagator has a pair of complex conjugate poles or branch
points with timelike real parts and large imaginary parts.  Such a propagator
does not have a Lehmann representation and admits the interpretation of
describing a confined particle.

Our study illustrates the manner in which the DSEs can be used to develop a
semi-phenomenological approach to QCD that incorporates the perturbative,
large spacelike-$k^2$ behaviour known from renormalisation group studies and,
via an economical parametrisation, extrapolates this into the
nonperturbative, small spacelike-$k^2$ domain.  This efficacious,
nonperturbative approach allows for the correlation of a large range of
observables via very few parameters, which it may be possible to relate to
the fundamental parameters of QCD.

\acknowledgements
The preparation of this manuscript has benefited from useful discussions with
A. Bender, F. T. Hawes and A. G. Williams.  This work was supported by the US
Department of Energy, Nuclear Physics Division; that of MRF under contract
No. DE-FG06-90ER40561 and that of CDR under contract No. W-31-109-ENG-38.
The calculations described herein were carried out using a grant of computer
time and the resources of the National Energy Research Supercomputer Center.



\begin{table}
\begin{tabular}{|l|l|l|}
        & Calculated    & Experiment \\ \hline
$m_\pi^{\rm Mass-Formula:}[B_0]$ & 138.7 MeV       & 138.3$\pm$ 0.5 \\ \hline
$m_\pi^{\rm Mass-Formula:}[F_0]$  & 137.2       &   \\ \hline
$m_\pi^{\rm BS-Equation}$  & 139.5     & \\ \hline\hline
$f_\pi[F_0]$      & ~92.4 MeV     & ~92.4$\pm$ 0.3 \\ \hline
$f_\pi[B_{0}]$    & ~92.3    &       \\ \hline
$N_\pi[F_0]$ & 102      & \\ \hline\hline
$r_\pi[F_0] N_\pi[F_0]$  & ~0.24          & ~0.31$\pm$ 0.004 \\ \hline
$a_0^0[F_0]$ & ~0.16          & ~0.21$\pm$ 0.02 \\ \hline
$a_0^2[F_0]$ & -0.041         & -0.040 $\pm$ 0.003\\ \hline
$a_1^1[F_0]$ & ~0.028         & ~0.038 $\pm$ 0.003\\ \hline
$a_2^0[F_0]$ & ~0.0022        & ~0.0017 $\pm$ 0.0003 \\ \hline
$a_2^2[F_0]$ & ~0.0013        & \\ \hline\hline
$g_{\pi^0\gamma\gamma}[F_0]$    & ~0.45          & ~0.50 $\pm$ 0.02 \\ \hline
$m_\rho[F_0^\rho]$ & ~0.971~GeV  & ~0.770 $\pm$ 0.001 \\ \hline
$g_{\rho\pi\pi}[F_0^\rho]$ & ~4.07 & ~6.07 $\pm$ 0.02 \\
\end{tabular}
\caption{Observables calculated using the parameter values in
Eq.~(\protect\ref{params}). The experimental values of the $\pi$-$\pi$
scattering lengths are discussed in Refs.~\protect\cite{RCSI94,S92}.  The
other experimental values are are taken from Ref.~\protect\cite{PDG94}.
$[B_{0}]$ indicates that the quantity was calculated using the approximation
of Eq.~(\protect\ref{goldthm}) while $[F_0]$ indicates it was calculated
using the zeroth order Tschebyshev moment obtained in a direct solution of
the BSE, Sec.~\protect\ref{secbse}.  The anomalous coupling
$g_{\pi^0\gamma\gamma}$ is discussed in Ref.~\protect\cite{CDRpion}.  See
Sec.~\protect\ref{secrho} for a discussion of the $\rho$-meson observables.
The difference between $N_\pi$ and $f_\pi$ is a measure of the accuracy of
the approximation of Eq.~(\protect\ref{pileading}).  That between the
calculated and experimental values of $r_\pi N_\pi$ is a measure of the
importance of final-state $\pi$-$\pi$ interactions and photon-$\rho$-meson
mixing.\protect\cite{ABR95} Final-state $\pi$-$\pi$ interactions are also
neglected in the calculation of the scattering lengths\protect\cite{RCSI94}
and $g_{\rho\pi\pi}$.  Pion-loop corrections to $m_\rho$ are of the order of
5\%.\protect\cite{HRM92,TM95}
\label{tabobs}}
\end{table}


\begin{references}
\bibitem{Euclidean} Our Euclidean metric conventions are
\mbox{$a\cdot b = \sum_{\mu,\nu=1}^4\,\delta_{\mu\nu}\,a_\mu\,b_\nu$}, with
\mbox{$\delta_{\mu\nu} = {\rm diag}(1,1,1,1)$} the Kronecker-delta.  Our
Euclidean Dirac matrices are Hermitian and satisfy
\mbox{$\{\gamma_\mu,\gamma_\nu\}=2\delta_{\mu\nu}$}.  A vector, $k$, is
spacelike if $k^2>0$.
\end{references}
\end{document}